\documentstyle[preprint,aps]{revtex}
\newcommand{\bmp}{{\mbox{\boldmath $p$}}}
\newcommand{\bms}{{\mbox{\boldmath $s$}}}

\newcommand{\bmr}{{\mbox{\boldmath $r$}}}
\newcommand{\bmz}{{\mbox{\boldmath $z$}}}

\newcommand{\bmq}{{\mbox{\boldmath $q$}}}

\begin{document}
\preprint {WIS-96/26/Jun-PH}
\draft

\date{\today}
\title{Comment on Final State Interactions beyond the Plane Wave Impulse
Approximation description of inclusive scattering}
\author{A.S. Rinat}
\address{Department of Particle Physics, Weizmann Institute of
         Science, Rehovot 76100, Israel}
\maketitle
\begin{abstract}

It is shown that no set of reasonable approximations leads to a
structure function as a convolution of the PWIA and a FSI contribution.

\end{abstract}
\medskip
Benhar $et\,  al$ have proposed for the structure
function of a composite system a form based on the Plane Wave Impulse
Approximation (PWIA), corrected for Final State  Interaction (FSI)
\cite{ben}. Although the result has been critically reviewed before
\cite{asr} a  recent preprint \cite{ben3} takes much the same
stand and we renew an attempt to clarify the issue.

Consider the dominant incoherent part of the structure function per
particle due to a density fluctuation
$\rho_q(\bmr_1)={\rm exp}(i\bmq\bmr_1)$ of a selected particle $'$1$'$
\begin{eqnarray}
S(q,\omega)= (m/q)\phi(q,y)=-(1/\pi){\rm Im} \bigg\langle \Phi_A^0|
\rho_q^{\dagger}(\bmr_1) G(\omega+E_A^0+i\epsilon)\rho_q(\bmr_1)
|\Phi_A^0\bigg\rangle
\label{a1}
\end{eqnarray}
$G(z)=(z-H_A)^{-1}$ is the Green's function of the system.
$\bmq,\omega$ are the momentum and energy transfer  in an inclusive
reaction from which $S$ is extracted and $\phi(q,y)$  is the reduced
response with the energy loss $\omega$ replaced by
$y=(m/q)(\omega-q^2/2m)$ \cite{grs,west}. Benhar et al  proposed
\cite{ben,ben3}
\begin{mathletters}
\label{a3}
\begin{eqnarray}
\phi_B(q,y)=(q/m)S_B(q,\omega)
&=&{\rm Re}\int dy' \phi_{PWIA}(q,y-y')R_B(q,y')
\label{a3a}\\
{\tilde R}_B(q,s)=\int dy e^{-iys} R_B(q,y)&=&
{\rm exp}\bigg \lbrack  \int d\bmr \rho_2(\bmr,0;\bmr,0)
{\rm exp}\bigg (i{\tilde \chi}_1(q,\bmr;s)\bigg )-1\bigg \rbrack
\label{a3b}\\
&\to & {\rm exp}\bigg \lbrack i \int d\bmr \rho_2(\bmr,0;\bmr,0)
{\tilde \chi}_1(q,\bmr;s)\bigg \rbrack
\label{a3c}
\end{eqnarray}
\end{mathletters}
Here
\begin{eqnarray}
(m/q)\phi_{PWIA}=S_{PWIA}(q,\omega)=\int d\bmp
P\bigg (\bmp,\omega-\frac {(\bmp+\bmq)^2}{2m}\bigg ),
\label{a4}
\end{eqnarray}
is the response  without core recoil in  the PWIA in terms  of the single
particle spectral   function   $P$.   $R_B$   accounts   for   FSI and
is expressed by means of the diagonal two-particle density matrix and
an eikonal phase
$\tilde{\chi}_1(q,\bmr,s)=(m/q)\int_0^sds'V(\bmr-s\hat\bmq)$. It is
totally off-shell in coordinate space with finite integration limits
instead of $-\infty,\infty$, and reflects a
particle, knocked-out inside the medium which will ultimately
not be detected.    Finally  Eq. (\ref{a3c})  is   the  Born
approximation of (\ref{a3b}) for weak $V$.

We now attempt to derive (\ref{a3}) within a framework,  suitable for
the discussion of approximations.
It suffices to consider a non-relativistic, infinite system
with Hamiltonian,  states and  energies $H_A=T+V, \Phi_A^n$  and $E^n_A$.
Assuming local forces
\begin{mathletters}
\label{a5}
\begin{eqnarray}
\rho^{\dagger}_q(\bmr_1)H(\bmp_n,\bmr_n)\rho_q(\bmr_1)&=&
H(\bmp_n,\bmr_n)+{\cal L}(\bmp_1,\bmq)
\label{a5a}\\
{\cal L}(\bmp_1,\bmq)&=&(\bmp_1+\bmq)^2/2m-\bmp_1^2/2m=
\bmq^2/2m+p_{1z}q/m,
\label{a5b}
\end{eqnarray}
\end{mathletters}
where $\hat\bmq=\hat \bmz$. With a correspondingly shifted Green's
function $\Gamma(z)=\rho_q^{\dagger}(\bmr_1)G(z)\rho_q(\bmr_1)$
(\ref{a1}) becomes \cite{grs}
\begin{mathletters}
\label{a7}
\begin{eqnarray}
S(q,\omega)
&=&-(1/\pi){\rm Im}\bigg\langle
\Phi^0_A|\Gamma(\omega+E_A^0+i\epsilon)|\Phi^0_A\bigg\rangle
\label{a7a}\\
\Gamma&=&\bigg (\omega+E_A^0-H_A(\bmr_1;\bmr_j)+{\cal L}(\bmp_1,\bmq)
+i\epsilon\bigg )^{-1}
\label{a7b}\\
&=&\bigg (\omega+E_A^0-H_{A-1}(\bmr_j)+T_{\bmp_1}+V_1(\bmr_1,\bmr_j)
+{\cal L}(\bmp_1,\bmq)+i\epsilon\bigg )^{-1},
\label{a7c}
\end{eqnarray}
\end{mathletters}
with $V_1(\bmr_1;\bmr_j)=
\Sigma_{j\le 2} V(\bmr_1-\bmr_j)$, the residual
interaction of $'$1$'$ with the core.

In practice one computes the  structure  function $S$,
retaining  parts in the  argument  of $\Gamma$ and disregarding  the
remainder $\gamma$.  The corresponding response is then
\begin{eqnarray}
S=S_d+\Delta S(\gamma_d)\approx S_d=S_{d,0}+S_{d,FSI},
\label{a8}
\end{eqnarray}
with some leading contribution $S_{d,0}$ and a
Final State Interaction (FSA) part $S_{d,FSI}$.
We discuss two choices
\begin{mathletters}
\label{a9}
\begin{eqnarray}
\Gamma_a&=&\bigg (\omega+{\cal L}(\bmp_1,\bmq)-V_1(\bmr_1;\bmr_j)
+i\epsilon\bigg )^{-1},
\,\,\,\, \gamma_a=E_A^0-H_A+V_1(\bmr_1;\bmr_j)
=E_A^0-H_{A-1}-T_{\bmp_1}
\label{a9a}\\
\Gamma_b&=&\bigg (\omega+E_A^0-H_{A-1}(j)+T_{\bmp_1+\bmq}
+i\epsilon\bigg )^{-1},\,\,\,\, \gamma_b=V_1(\bmr_1;\bmr_j),
\label{a9b}
\end{eqnarray}
\end{mathletters}

In a) the kinetic energy in $H_A$ resides in the neglected part
$\gamma_a$ and thus suits a large-$q$ approximation, where after
absorbing $\bmq$, the recoiling  particle   $'$1$'$ has momentum
$\,\,|\bmp_1+\bmq|\approx q\gg p_1,p_j$.  Its kinetic energy in the
appropriate eikonal approximation is ${\cal L}$, Eq. (\ref{a5b}).
For the reduced response one has for weak $V$ \cite{grs}
\begin{mathletters}
\label{a10}
\begin{eqnarray}
\phi_a(q,y)&=& \frac{1}{2\pi}{\rm Re} \int_{-\infty}^{\infty} ds
e^{isy}\bigg \lbrack \frac{\rho_1(0,s)}{\rho}+ i\int d\bmr
\frac{\rho_2(\bmr,0;\bmr-s\hat\bmq,0)}{\rho}
{\tilde \chi}(q,\bmr;s)+... \bigg\rbrack
\label{a10a}\\
{\tilde\chi}={\tilde\chi}_1+{\tilde\chi}_2
&=&(m/q)\bigg \lbrack \int_0^s ds'
V(\bmr-s'\hat \bmq)-sV(\bmr-s\hat \bmq)\bigg \rbrack
\label{a10b}
\end{eqnarray}
\end{mathletters}
Eqs. (\ref{a10}) give the lowest order terms in a $1/q$
expansion of $\phi$ in terms of non-diagonal density matrices
$\rho_1,\rho_2$. In particular the asymptotic limit ($\bms=s\hat \bmq$)
\begin{eqnarray}
F_0(y)=\lim_{q \to \infty}\phi(q,y)=
(4\pi)^{-2}\int_{|y|}^{\infty} dp p n(p)=
(4\pi)^{-2}\int_{|y|}^{\infty} dp p
\int d\bms\,\, {\rm exp}[i\bms\bmp](\rho_1(0,s)/\rho)
\label{a11}
\end{eqnarray}
The first cumulant corresponding to (\ref{a10}) is \cite{grs2}
\begin{mathletters}
\label{a12}
\begin{eqnarray}
\phi_a(q,y)&=&\frac{1}{2\pi}{\rm Re}\int ds e^{isy}
\frac {\rho_1(0,s)}{\rho} \tilde R_a(q,s)=
{\rm Re} \int dy' F_0(y-y')R_a(q,y')
\label{a12a}\\
{\tilde R}_a(q,s)&=&
{\rm exp}\bigg \lbrack \int d\bmr
\frac{\rho_2(\bmr,0;\bmr-s\hat\bmq,0)} {\rho_1(0,s)}
\bigg (e^{i\tilde{\chi}(q,\bmr;s)}-1 \bigg ) \bigg\rbrack
\to {\rm exp}\bigg \lbrack i\int d\bmr
\frac{\rho_2(\bmr,0;\bmr-s\hat\bmq,0)}
{\rho_1(0,s)}{\tilde\chi}(q,\bmr;s)\bigg \rbrack
\label{a12b}
\end{eqnarray}
\end{mathletters}

In case b) one starts from  the exact shifted Green's function except for
the   residual  interaction   $V_1$.    Insertion  of   a  complete   set
$\Phi_{A-1}^n$ into (\ref{a7a}) leads to the PWIA
\begin{eqnarray}
S_b(q,\omega)=S_{PWIA}(q,\omega)=\int d\bmp
P\bigg (\bmp,\omega-\frac {(\bmp+\bmq)^2}{2m}\bigg ),
\label{a13}
\end{eqnarray}
with $P(p,E)$ the single-particle spectral function of the target.

According to (\ref{a8}), the above is the dominant part with no
FSI left, unless one considers
the otherwise neglected $\gamma_b=V_1(\bmr_1;\bmr_j)$. Inclusion of
generally, non-diagonal core matrix-elements of $V_1$ causes grave
complications. It has been suggested to replace
$V_1(\bmr_1;\bmr_j)$ by an optical potential ${\cal V}^{opt}$
\begin{eqnarray}
\langle\bmp,n|\gamma_b|\bmp',n'\rangle
\to \langle\bmp,0|V_1(i;j)|\bmp',0\rangle\to
\langle\bmp|{\cal V}^{opt}|\bmp'\rangle
\label{a14}
\end{eqnarray}
However, by definition ${\cal V}^{opt}$ replaces only $diagonal$ matrix
elements of $V_1(\bmr_1;\bmr_j)$ and the approximation (\ref{a14}) is
consequently  impermissible.

With a fast recoiling particle also in b) one is tempted to introduce
a Fixed Scatterers Approximation (FSA) for $V_1$. This, however,
seems not commensurate with the retention of a dynamically active
$H_{A-1}$. The objection is circumvented if
$E_A^0-H_{A-1}(j)$ in (\ref{a7b}) is  replaced
by an average separation energy $\langle \Delta\rangle$, or
equivalently, if closure is applied to the states $\Phi_n^{A-1}$,
implicit in $P$, Eq. (\ref{a13}).
Neglecting in addition  $\bmp_1^2\ll {\cal L}$ one finds
\begin{eqnarray}
\phi_b^{clos}(q,y)\to \phi_a(q,y-\langle \Delta\rangle)
\label{a15}
\end{eqnarray}
Except for a small shift, relevant only around the quasi-elastic peak
$y=0$, closure reduces case b) to a).

We return to the expression (\ref{a3}) of Benhar  $et\, al.$
\cite{ben,ben3}, passing over  many  intermediate heuristic  steps  in
its  construction.
Its  form resembles
(\ref{a12}) which without proof has been assumed to also
hold if  the leading  asymptotic part $S_{as} (=F_0)\to S_{PWIA}$.
The discussion of case b) shows this not to be possible, even
when assuming (\ref{a14}).

Next  the authors  claim (\ref{a3})  to  be the  same as  used by  Silver
\cite{sil}, itself a  re-derivation  of results  in \cite{grs,grs2},  but
there  are obvious  differences.   We mention  the  diagonal $\rho_2$  in
(\ref{a3}), different from the semi-diagonal in (\ref{a12b}) and also one
potential term out  of the two in (\ref{a10b}).  The  second one vanishes
only for hard-core interactions, using a non-diagonal $\rho_2$.

Putting aside the derivation of (\ref{a3}) we next assess the actual
difference, comparing $1/q$  expansions of $\phi_a,\phi_{PWIA}$, which
can be shown to have the same asymptotic limit $F_0(y)$,
Eq. ({\ref{a11}). From  Eqs. (\ref{a10}) and
(\ref{a3}) one finds  for the Fourier transforms  of
the lowest order FSI term $\phi_1(q,y)=(m/q)F_1(y)$
\begin{mathletters}
\label{a16}
\begin{eqnarray}
{\tilde F}_{a,1}(s)&=&
+i\int d\bmr\frac{\rho_2(\bmr,0;\bmr-s\hat\bmq,0)}{\rho}
\bigg \lbrack \int_0^s ds'
V(\bmr-s'\hat \bmq)-sV(\bmr-s\hat \bmq)\bigg \rbrack
\label{a16a}\\
{\tilde F}_{PWI ,1}(s)  &=&
\Sigma_n\int d\bmp \, e^{ip_zs}\,|\Gamma_n(p)|^2
\bigg (\frac{p^2}{2m}-\Delta_n\bigg )
+i \int d\bmr\frac{\rho_2(\bmr,0;\bmr,0)}{\rho}
\int_0^s ds'V(\bmr-s'\hat \bmq)
\nonumber\\
&=& i \int d\bmr\bigg \lbrack
\frac{\rho_2(\bmr,0;\bmr,0)}{\rho} \int_0^s ds'V(\bmr-s'\hat \bmq)-
\frac{\rho_2(\bmr,0;\bmr-s\hat\bmq,0)}{\rho}
s V(\bmr-s\hat \bmq)\bigg \rbrack
\label{a16b}
\end{eqnarray}
\end{mathletters}
The  proof leading to  the second  equation in (\ref{a16b})  can be
found in \cite{rd}.   One notices diagonal and  semi-diagonal $\rho_2$ as
opposed to  exclusively the half-diagonal  one in (\ref{a16a}).   The two
expressions  (\ref{a16}) thus  differ and  so do  their first  cumulants.
Numerical consequences have  been discussed in a  comparison of inclusive
cross section  of electrons  from Fe and  Nuclear Matter  \cite{asr}.  In
particular  for low  energy losses  there are considerable
differences in  the relative magnitude of  the leading and FSI  parts, as
well as in the total results for (\ref{a12}).

We  return to  the  simplifying assumptions made in the introduction:
Should short-range repulsion  in $V$  produce large integrals in (10c),
their  contribution  may be  tempered  by  a re-summation  $V\to
V_{eff}=t$  \cite{as} (cf. Eq. (\ref{a3b})). Also,
the above  presentation  is  not
different  for finite  targets,  nor is  it  qualitatively modified  when
minimal relativistic effects \cite{ben,asr} are applied.

We conclude that, starting from a  well-defined theory, there seems to be
no way to actually derive the  total response in a form (\ref{a3}) which
modifies a leading PWIA by FSI terms.

\bigskip

I thank S.A. Gurvitz  for useful comments and suggestions.

\end{document}